# BRIDGE CONFIGURATIONS IN PIEZORESISTIVE TWO-AXIS ACCELEROMETERS

*Einar Halvorsen and Svein Husa*

Institute of microsystems technology,
Faculty of Science and Engineering,
Vestfold University College,
P.O. Box 2243, N-3103 Tønsberg, Norway

**ABSTRACT**

In piezoresistive two-axis accelerometers with two proof masses suspended by cantilever beams, there are generally many ways to configure the Wheatstone bridges. The configurations are different both with respect to functionality and performance. The main distinction is between bridges that contain resistors belonging to both proof masses, and the one bridge that doesn't.

We compare the different bridge configurations by analytical calculations of bridge non-linearity, robustness towards manufacturing variations and electronic noise. We consider accelerometers where the ratio between the sensitivity to acceleration normal and parallel to the chip plane vary over a wide range. For numerical examples we use representative values for p-type silicon.

The performance of the configuration with one bridge connected to each proof mass is superior to those that combine resistors belonging to different proof masses.

## 1. INTRODUCTION

In symmetric two-axis accelerometers based on two proof masses supported by cantilever beams with piezoresistive sensing elements, there are several ways to configure the Wheatstone bridges [1-3]. The choice is between one bridge per proof mass and a variety of configurations combining piezoresistors that belong to different proof masses.

Resistors from different proof masses can be combined into one bridge in much the same way as unwanted cross-axis sensitivity is eliminated in single-axis devices [4]. This makes it possible to have electrical outputs that correspond to accelerations in mutually orthogonal directions, parallel or orthogonal respectively, to the chip plane [2]. If this is the form of output that is desired, it can be hardwired into the system without any need for further post-processing.

Another potential advantage is that the bridge for motion parallel to the chip plane, can be placed anywhere on the beam resulting in larger degree of freedom in the chip-layout and possibly also smaller size.

In piezoresistive sensing, the dominating noise source is in most cases electronic noise from the bridge resistors, in particular at low frequencies where 1/f noise dominates [5]. The noise voltage has a fixed dependence on the excitation voltage, $V_{ex}$, for a given resistor layout. The noise is therefore the same from all bridges regardless of the configuration, while the useful signal is smaller when cancellations occur in the bridge. Consequently, bridge configurations that use resistors from different proof masses have a clear disadvantage from a signal to noise ratio perspective. The question then arises if these configurations have other performance measures, such as linearity and tolerance to manufacturing variations, that still give them advantages?

Depending on the design choices or process constraints, the sensitivity (scale factor) of cantilever accelerometers may differ greatly between the in-plane acceleration and the out of plane acceleration. This property can be characterised by a sensitivity angle which gives the direction of sensitivity for a cantilever/proof mass subsystem. In 3-axis accelerometers the sensitivity angle is known to influence the overall noise limited resolution of the device [6]. It is therefore interesting to look into how sensitivity angle affects the noise in the measured acceleration.

In the following sections we investigate by detailed analytical calculations four resistor bridge configurations for two-axis accelerometers. They are analysed with respect to the issues raised above. In Section 2 we discuss the possible configurations of the bridges. Section 3 compares bridge nonlinearities as a function of sensitivity angle. Section 4 discusses tolerance to deviations in local stress. Section 5 investigates tolerances to variations in the resistances values. Noise in the measured acceleration is analysed in Section 6 both with respect to the direction of the acceleration and the sensitivity angle of the device. A final summary and conclusions are found in Section 7.

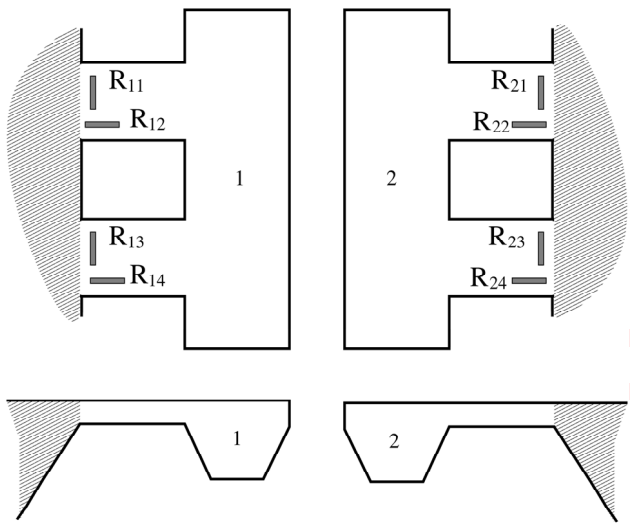

**Figure 1** Two-axis accelerometer structure with rough indication of piezoresistor positions. Upper figure: viewed from top. Lower figure: cross section.

## 2. BRIDGE CONFIGURATIONS

We first classify bridge configurations using the simplified picture that the resistance change of longitudinally and transversally oriented resistors are equal in magnitude and opposite in sign.

We consider bridge configurations with reference to the accelerometer structure sketched in Figure 1. The schematics of all configurations are shown in Figure 2. The configuration where all resistors belonging to the same proof mass are in the same bridge is named configuration A. Next, we name configuration B, the configuration that uses the same resistors as above, but with resistors from both proof masses in each voltage divider.

It is also possible to make a pair of bridges where in each bridge, the resistors that belong to the same proof mass are equally oriented. We can use, say, six longitudinally and two transversally oriented resistors. We consider the case when, in Figure 1, $R_{11}$ and $R_{13}$ are substituted by longitudinally oriented resistors $R'_{11}$ and $R'_{13}$ respectively. The configuration is named C in Figure 2.

We finally consider a configuration D in which each voltage divider in both bridges has resistors from the same proof mass. This configuration has a substantial common mode signal, but is included for completeness.

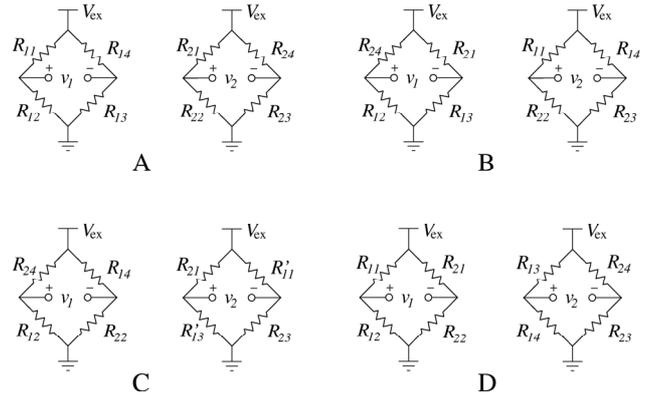

**Figure 2** Bridge configurations.

In analysing the system, it is convenient to introduce the column matrices,

$$a = \begin{bmatrix} a_1 \\ a_2 \end{bmatrix}, \ \sigma = \begin{bmatrix} \sigma_1 \\ \sigma_2 \end{bmatrix}, \ v = \begin{bmatrix} v_1 \\ v_2 \end{bmatrix} \quad (1)$$

which contain respectively the acceleration in directions 1 and 2, the principal stress at the base of the beams for proof mass 1 and 2 and the output voltages from bridges 1 and 2.

In the linear regime, the stress in the beam is linearly related to the acceleration through $\sigma = Ha$ where $H$ is a two by two matrix. For a perfectly symmetric accelerometer $H$ takes the value

$$H_0 = h \begin{bmatrix} \cos\alpha & \sin\alpha \\ -\cos\alpha & \sin\alpha \end{bmatrix} \quad (2)$$

where $h$ is a constant and $\alpha$ is the angle of sensitivity.

The output from the bridges can be written

$$v = v_0 + Ta + O(a^2) + v_n \quad (3)$$

where $v_0$ contains the offset voltages, $v_n$ contains the resistor bridge noise voltages and $T$ is a two by two matrix parametrising the linear dependence of the output voltages on the acceleration. There are also nonlinear terms for which we do not define any notation.

Define a matrix $W$ such that

$$v = W\sigma. \quad (4)$$

Then the nominal value of $T$ is

$$T_0 = WH_0. \quad (5)$$

$W$ is dependent on the bridge configuration. $H_0$ is not.

In an actual device, a longitudinal piezoresistor that belongs to proof mass $i$, will have a resistance $R = R_0(1 + \pi_l \sigma_i)$ and a transversal piezoresistor will have a resistance $R = R_0(1 + \pi_t \sigma_i)$. $R_0$ is the resistance in the absence of a stress and is chosen by the design. The piezoresistance coefficients $\pi_l$ and $\pi_t$ are material parameters. In numerical examples, we will use representative values for p-type Si: $\pi_l = 71.8 \cdot 10^{-11}$ Pa$^{-1}$ and $\pi_t = -66.3 \cdot 10^{-11}$ Pa$^{-1}$ [7]. When we expand the output voltage to first order in the stresses for each of the four configurations, we get the following results:

$$W^A = \frac{V_{ex}}{2}\begin{bmatrix} \pi_l - \pi_t & 0 \\ 0 & \pi_l - \pi_t \end{bmatrix}, \quad (6)$$

$$W^B = \frac{V_{ex}}{4}\begin{bmatrix} \pi_l - \pi_t & -\pi_l + \pi_t \\ \pi_l - \pi_t & \pi_l - \pi_t \end{bmatrix}, \quad (7)$$

$$W^C = \frac{V_{ex}}{2}\begin{bmatrix} \pi_l & -\pi_l \\ \pi_l & -\pi_t \end{bmatrix}, \quad (8)$$

$$W^D = W^B. \quad (9)$$

The measured or estimated acceleration is

$$\tilde{a} = T_0^{-1} v, \quad (10)$$

and the error is

$$\delta a = \tilde{a} - a = T_0^{-1} v - a = \Delta a + a_n. \quad (11)$$

On the right hand side we have decomposed the error into a noise part,

$$a_n = T_0^{-1} v_n \quad (12)$$

and a non-random part, $\Delta a$.

In the following we analyse the different contributions to the measured error one by one and compare the performance of the four bridge configurations.

### 3. BRIDGE NONLINEARITY

If we expand the bridge output voltages (3) to second order in the acceleration, we find that the leading corrections to linear behavior are :

$$\Delta a^A = \Delta a^B = \Delta a^D = -\frac{h\|a\|^2}{4}(\pi_l + \pi_t)$$
$$\times \begin{bmatrix} \frac{\cos^2(\theta - \alpha) - \cos^2(\theta + \alpha)}{\cos \alpha} \\ \frac{\cos^2(\theta - \alpha) + \cos^2(\theta + \alpha)}{\sin \alpha} \end{bmatrix} \quad (13)$$

and

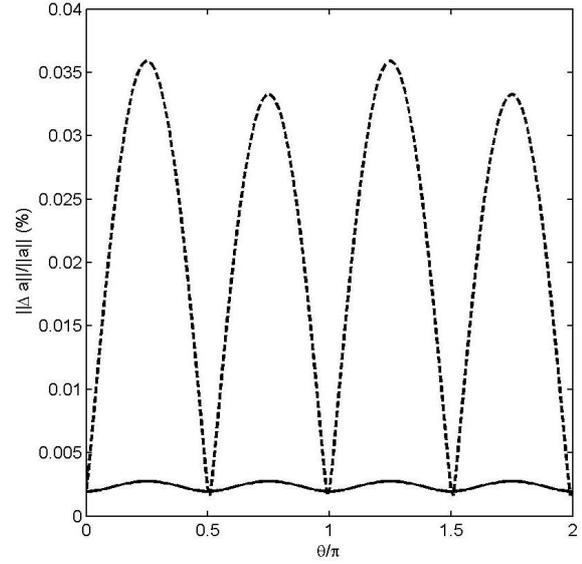

**Figure 3** Bridge nonlinearity contribution to error in estimated acceleration for α=π/4, *h*=1MPa/*g*, *a*=1*g* and p-type silicon piezoresistors. Solid line: configurations A, B and D. Dashed line: configuration C.

$$\Delta a^C = -\frac{h\|a\|^2}{4}$$
$$\times \begin{bmatrix} \pi_l \frac{\cos^2(\theta - \alpha) - \cos^2(\theta + \alpha)}{\cos \alpha} \\ \frac{\pi_l \cos^2(\theta - \alpha) + (\pi_l + 2\pi_t)\cos^2(\theta + \alpha)}{\sin \alpha} \end{bmatrix}. \quad (14)$$

Here we have introduced polar coordinates for the acceleration: $a_1 = \|a\|\cos\theta$ and $a_2 = \|a\|\sin\theta$.

Figure 3 shows the relative magnitude of the nonlinearity error for the four configurations for realistic values of acceleration and sensitivity. We see that the nonlinearity error depends on orientation and that judging from worst case over the range of $\theta$, configurations A, B and D are requal to each other while C is much worse than the others.

The maximum nonlinearity error over range of $\theta$ is plotted vs. the sensitivity angle in Figure 4. Clearly C has the worst performance.

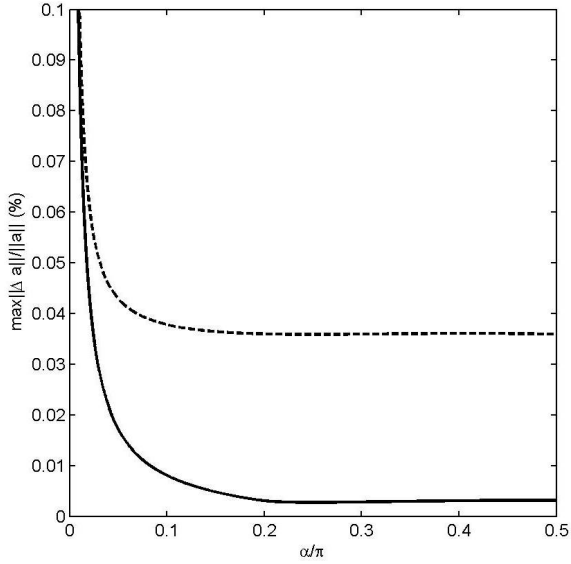

**Figure 4** Maximum bridge nonlinearity error in estimated acceleration vs. sensitivity angle for $h$=1MPa/$g$, $a$=1$g$ and p-type silicon piezoresistors. Solid line: configurations A, B and D. Dashed line: configuration C.

### 4. DEVIATIONS IN STRESS

It is possible that process variations result in deviations from nominal values in the stress at piezoresistors. The deviations can be described by a change $\Delta H$ in the linear relation between stress and the acceleration of the package frame with respect to the nominal value $H_0$. The error in the acceleration is then

$$\Delta a = H_0^{-1}\Delta H a, \quad (15)$$

which is independent of the bridge configuration. The four configurations are therefore equally robust against deviations in the mechanical structure.

### 5. OFFSET SIGNAL

There may be deviations from nominal in the value of the piezoresistors. If these deviations are different between some or all resistors in a bridge, it results in a nonzero offset signal. In general we must consider correlations between all eight resistors, but we will be content with considering only a few special cases.

The resistance $R_I$ of any particular resistor labelled $I$ can be written $R_I = R_{I0}(1+\Delta_I)$ where $R_{I0}$ is the nominal resistance and $\Delta_I$ is the relative deviation from nominal which we will treat as a random variable with zero mean and variance $\Delta^2$. The values of the different resistors will be taken to be either identical or statistically independent.

We consider only small variations in resistivity so that we can work with effects that are linear in the deviations from nominal. We can express the offset signal as $\Delta a = T^{-1}v_0$, where the offset voltage, $v_0$, is found by developing the different expressions to first order in all $\Delta_I$. As a measure of the offset error we use the expectation of the square magnitude of the offset error $\langle \|\Delta a\|^2 \rangle$. We will investigate three cases.

First we consider all deviations statistically independent. Following the procedure outlined above, we find

$$\langle \|\Delta a_A\|^2 \rangle = \frac{2\Delta^2}{[h(\pi_l-\pi_t)\sin 2\alpha]^2} \quad (16)$$

$$\langle \|\Delta a_B\|^2 \rangle = \langle \|\Delta a_D\|^2 \rangle = 2\langle \|\Delta a_A\|^2 \rangle \quad (17)$$

$$\langle \|\Delta a_C\|^2 \rangle = \frac{3\pi_l^2 + \pi_t^2 + 2\pi_l(\pi_l+\pi_t)\cos 2\alpha}{4\pi_l^2} \langle \|\Delta a_B\|^2 \rangle \quad (18)$$

It is seen that the squared error is twice as big for configuration B and D as for configuration A. Configuration C has the same error as for configuration B when the piezoresistance coefficients are opposite in sign and equal in magnitude.

We next consider identical deviations for all resistors belonging to the same proof mass, but statistically independent for resistors belonging to different proof masses. The result is

$$\langle \|\Delta a_A\|^2 \rangle = \langle \|\Delta a_B\|^2 \rangle = \langle \|\Delta a_D\|^2 \rangle = 0, \quad (19)$$

$$\langle \|\Delta a_C\|^2 \rangle = \frac{4\Delta^2}{[h\pi_l \sin 2\alpha]^2}. \quad (20)$$

We finally consider statistically independent deviations between longitudinally and transversally oriented resistors, but identical deviations within each of the two groups, regardless of location. In this case, they are all equal:

$$\langle \|\Delta a_A\|^2 \rangle = \langle \|\Delta a_B\|^2 \rangle = \langle \|\Delta a_C\|^2 \rangle = \langle \|\Delta a_D\|^2 \rangle$$
$$= \frac{2\Delta^2}{[h(\pi_l-\pi_t)\sin\alpha]^2} \quad (21)$$

## 6. NOISE

All resistors have the same nominal value and are made in the same process, so the resistor noise voltages all have the same power spectral density $S_R(\omega)$. It has contributions from Johnson noise and flicker noise. The noise voltage in each resistor is statistically independent of that of the other resistors. The noise in the estimated acceleration can then be written

$$S_a = S_R T_0^{-1} (T_0^{-1})^{\mathrm{T}}. \quad (22)$$

For the four configurations, the resulting power spectral densities are

$$S_a^A = \frac{2 S_R}{[h V_{\mathrm{ex}} (\pi_l - \pi_t)]^2} \begin{bmatrix} \dfrac{1}{\cos^2 \alpha} & 0 \\ 0 & \dfrac{1}{\sin^2 \alpha} \end{bmatrix}, \quad (23)$$

$$S_a^B = S_a^D = 2 S_a^A, \quad (24)$$

and

$$S_a^C = \frac{4 S_R}{[h V_{\mathrm{ex}} (\pi_l - \pi_t)]^2} \times$$

$$\begin{bmatrix} \dfrac{(\pi_l - \pi_t)^2}{4 \pi_l^2 \cos^2 \alpha} & -\dfrac{\pi_l^2 - \pi_t^2}{4 \pi_l^2 \sin \alpha \cos \alpha} \\ -\dfrac{\pi_l^2 - \pi_t^2}{4 \pi_l^2 \sin \alpha \cos \alpha} & \dfrac{4 \pi_l^2 - (\pi_l + \pi_t)^2}{4 \pi_l^2 \sin^2 \alpha} \end{bmatrix}. \quad (25)$$

The power spectral density of the noise in the estimated acceleration is twice as big for configuration B and D as for configuration A. Configuration C is more complicated, but equals that for configuration B when the piezoresistance coefficients are equal in magnitude and opposite in sign. Configuration C also has a nonzero cross spectral density.

The power spectral density of acceleration noise in a particular direction given by the angle $\theta$ with the x-axis can be found from

$$S(\theta) = n^{\mathrm{T}}(\theta) S_a n(\theta), \quad n(\theta) = \begin{bmatrix} \cos \theta \\ \sin \theta \end{bmatrix}. \quad (26)$$

For configuration A, this is

$$S^A(\theta) = \frac{2 S_R}{[h V_{\mathrm{ex}} (\pi_l - \pi_t)]^2} \left( \frac{\cos^2 \theta}{\cos^2 \alpha} + \frac{\sin^2 \theta}{\sin^2 \alpha} \right) \quad (27)$$

and therefore we have at the optimum sensitivity angle in the worst case direction:

$$S_0^A = \min_\alpha \max_\theta S^A(\theta) = \frac{4 S_R}{[h V_{\mathrm{ex}} (\pi_l - \pi_t)]^2}. \quad (28)$$

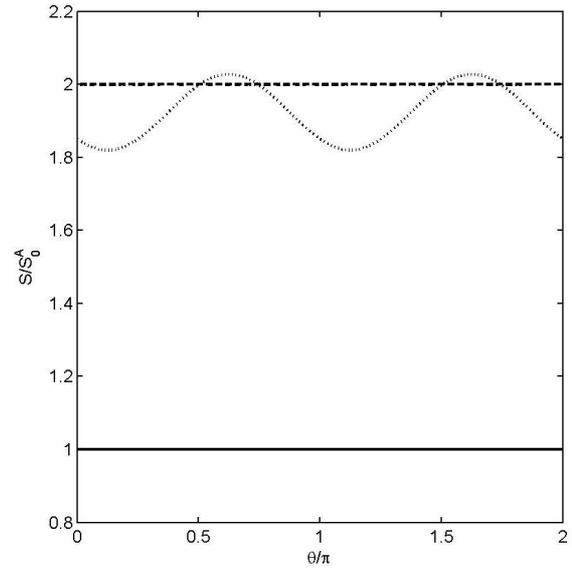

**Figure 5** Normalized noise power spectral densities for α=π/4. Solid line: configuration A, dashed line: configurations B and D, dotted line: configuration C.

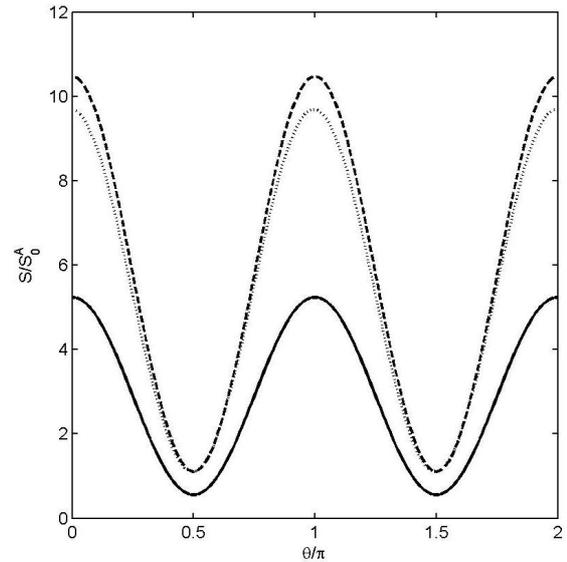

**Figure 6** Normalized noise power spectral densities for α=0.4π. Solid line: configuration A, dashed line: configurations B and D, dotted line: configuration C.

An example of the power spectral densities of the noise vs. direction with $\alpha = \pi/4$ is shown in Figure 5. For configurations A, B and D the noise level is independent of direction and related as commented above. For configuration C the noise level is weakly varying with direction from 9% below to 1.5% above the value for configurations B and D.

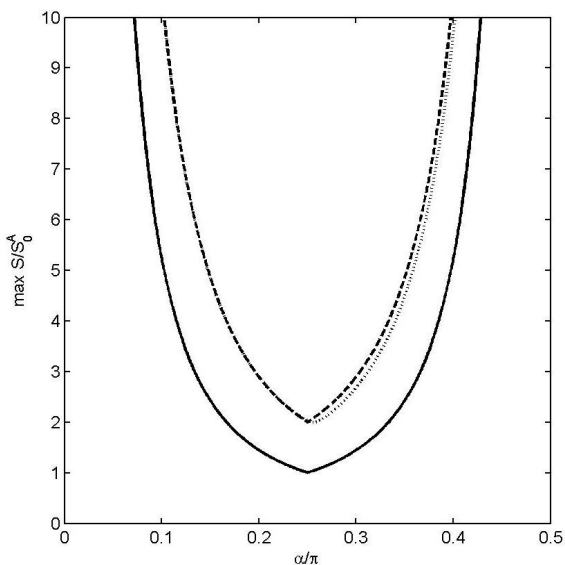

**Figure 7** Maximum normalized noise power spectral densities vs. α. Solid line: configuration A, dashed line: configurations B and D, dotted line: configuration C.

A second example is given in Figure 6. Here $\alpha = 0.4\pi$, which means that both proof masses are more sensitive to vertical accelerations than to horizontal accelerations. In the low sensitivity directions, noise influence increases, and the results in the figure show large noise in horizontal directions for all configurations. The best performance is for configuration A. The performance for configuration C is comparable to that of B and D.

In Figure 7 the worst case noise in any acceleration component is shown vs. the sensitivity direction. Again, A is best and C is comparable to B and D. The normalized noise diverges as the sensitivity direction approaches the singular cases in-plane or normal sensitivity directions.

## 7. SUMMARY AND CONCLUSION

We have considered a number of second order effects in two-axis accelerometers with differently configured resistor bridges.

For bridge nonlinearities we find that combination of resistors from both proof masses into one bridge, gives three configurations that are either equal to or much worse than having all resistors belonging to a proof mass in the same bridge.

We find that all configurations are equally sensitive to variations in the mechanical structure.

For deviations from the nominal value of the piezoresistors, we find that the spread in offset signal is in no case better than when the bridge has all resistors from the same proof mass.

The noise spectral density is about a factor 2 worse (angle dependent in one case) for the alternative cases compared to an accelerometer having one bridge for each proof mass.

In conclusion we have demonstrated that bridge configurations that use one bridge dedicated to each proof mass, is superior in performance to alternative bridge configuration. It can not be justified, merely based on performance considerations, to combine resistors belonging to different proof masses into each bridge.